\documentclass{article}
\usepackage[margin=1.0in]{geometry}
\usepackage[utf8]{inputenc}
\usepackage{lineno}
\providecommand{\keywords}[1]{\textbf{{Key-words:}} #1}
\usepackage{graphicx}                           
\usepackage{amsmath} 
\usepackage[font=footnotesize, labelfont=bf]{caption}
\usepackage{hyperref}
\hypersetup{colorlinks,linkcolor={black},citecolor={black},urlcolor={black}} 
\usepackage{color}
\usepackage{amsfonts}
\usepackage[semicolon]{natbib}
\usepackage{subcaption}
\usepackage{empheq}
\usepackage{dsfont}
\usepackage{lmodern}
\usepackage{array}
\usepackage{siunitx}
\usepackage{verbatim}
\usepackage{authblk}
\graphicspath{{images/}}

\graphicspath{{images/}}
\title{DyPy: A Python Library for Simulating Matrix-Form Games}
\author[1]{Anjalika Nande\footnote{Corresponding author. Email: anande01@g.harvard.edu}}
\author[3]{Andrew Ferdowsian}
\author[4]{Eric Lubin} 
\author[5]{Erez Yoeli}
\author[2]{Martin Nowak}
\affil[1]{Department of Physics, Harvard University}
\affil[2]{Departments of Mathematics and Organismic and Evolutionary Biology, Harvard University}
\affil[3]{Department of Economics, Princeton University}
\affil[4]{Department of Computer Science, Massachusetts Institute of Technology}
\affil[5]{Sloan School of Management, Massachusetts Institute of Technology}

\date{}
\begin{document}
\maketitle
\renewcommand{\abstractname}{Abstract}
\begin{abstract}
\noindent
Evolutionary Game Theory (EGT) simulations are used to model populations undergoing biological and cultural evolution in a range of fields, from biology to economics to linguistics. In this paper we present DyPy, an open source Python package that can perform evolutionary simulations for any matrix form game for three common evolutionary dynamics: Moran, Wright-Fisher and Replicator. We discuss the basic components of this package and illustrate how it can be used to run a variety of simulations. Our package allows a user to run such simulations fairly easily without much prior Python knowledge. We hope that this will be a great asset to researchers in a number of different fields.
\end{abstract}

\section*{Introduction}
Evolutionary Game Theory (EGT) is used to model populations undergoing Darwinian or cultural evolution; it has been used by researchers in biology, economics, sociology, anthropology, linguistics, and philosophy to explore topics such as multicellularity, human cooperation, and the evolution of grammar \citep{maynard1973,maynard1982,matsui1996,Traulsen2010,nowak2006evolutionary,gerhard2008,Bruin2005,Friedman1998,Johnson2003}.  While researchers can usually solve the evolutionary dynamics of simple games analytically, for many games, analytic solutions are difficult or impossible.  In such cases, researchers typically rely on numerical simulations to explore a game's dynamics \citep[for a review, see][]{nowak2006evolutionary}.  

In this paper we present DyPy, an open source Python package that can be used to easily run a wide range of useful EGT simulations which have up until now required duplicative and time-and-energy intensive coding on the part of researchers. The package provides functionality needed for most simple EGT simulations like permitting users to analyze any matrix-form games, regardless of the number of players and strategies, and regardless of whether the game is symmetric. Users can choose from three of the most commonly-used evolutionary dynamics: Moran, Wright-Fisher and Replicator. They can specify whether the population is subject only to individual-level selection, or whether there is also group-level selection. And, they can account for frequency biased imitation. Users can analyze a single run of the evolutionary dynamic, or many runs.  They can also vary model parameters along a range to explore the impact of the choice of parameter values on the dynamics of their game. The results of any simulation can be saved as a dataset containing the frequency of strategies over time and their payoffs.  These results can also be presented visually using a number of graphing functions.  Most simulations require just 20-100 lines of code, and can be written with little prior programming experience. DyPy has been parallelized and optimized to ensure simulations run fast. To the best of our knowledge, DyPy is unique in its scope and ease of implementation.

\section*{DyPy Availability} 
DyPy is an open source Python software library that is hosted on Github at \\ https://github.com/anjalika-nande/dynamics\_sim. It runs on Python 3.0 or higher. Detailed documentation for each command in the library, sample code for exemplary simulations and a Wiki is provided in the Github repository. Improvements through pull requests and suggestions for additional functionality are encouraged.

\section*{DyPy Output}
The result of each simulation is a data set consisting of the frequencies of strategies over time along with their associated payoffs. For ease of interpretation, these results can be presented visually using a number of different graphing options. These range from a 2D plot of the results of a single simulation to 3D wire and contour plots that are the result of varying two parameters over many iterations of a simulation (see Supporting Information for details about the 3D plotting options). All of these graphs can be saved in any file format supported by matplotlib, the default is .png.

\section*{Defining a Game}
There are two main components to an EGT simulation: the game that is being played and the dynamics through which strategies evolve. DyPy allows the user to create any desired game by subclassing the Game class and defining the payoff matrix appropriately. While creating the game, the user can also define states (i.e. the distribution of players playing each strategy) of interest--usually this is done to analyze possible equilibria of the game, which have been identified analytically. 

To illustrate how this is done and to showcase the library's functionality, we reproduce some of the well-known results from literature related to the evolution of cooperation throughout this paper. We start by defining the Prisoners' Dilemma.
\newline
\newline \textit{Prisoners' Dilemma}
\newline 

Cooperation is seen in biological systems at all scales; from the formation of multicellularity to large scale human cooperation \citep{Michod2001,Nowak2006}. However, cooperation which involves a personal cost for the benefit of others can always be exploited by defectors. This phenomenon is captured by the well-known two-player game, the Prisoners' Dilemma \citep{axelrod1980,maynard1982,nowak2006evolutionary}. The payoff matrix associated with the game is,

\begin{equation}
\bordermatrix{~ & C & D \cr
  C & R & S \cr
  D & T & P \cr}
\end{equation}

where $T>R>P>S$. 

Games like this are typically solved by finding the Nash equilibria of the game, which are strategy profiles such that no player can benefit by unilaterally deviating.  The only Nash equilibrium of this game is for both players to defect, even though if both cooperated, they would receive a higher payoff. 

In evolutionary game theory, there is an entire population of players, who, in each round, are assigned to play a game against each other--the details of how they are assigned to do so depend on the particular dynamic.  The most basic equilibrium concept in EGT is the Evolutionary Stable Strategy (ESS).  A strategy is an ESS if, when everyone in the population plays this strategy, a mutant who plays another strategy will receive lower payoffs and die out.  ESS corresponds closely to the Nash equilibrium (indeed, all ESS must be Nash equilibria, but not vice versa).  Just as the only Nash Equilibrium of the Prisoners' Dilemma is $(D,D)$, the only ESS of the Prisoners' Dilemma is for everyone in the population to play $D$ \citep{maynard1982}. 

We will now use DyPy to simulate a population of players playing the Prisoners' Dilemma to see that natural selection favors defection. To do this, we first set up the game. This involves specifying the strategies, equilibria and the payoff matrix associated with it. 
The following 16 lines of code are used to prepare the game:

\begin{verbatim}
    from games.game import SymmetricNPlayerGame
    
    # Class that defines the Prisoners' Dilemma game.
    class PrisonersDilemma(SymmetricNPlayerGame):
        DEFAULT_PARAMS = dict(R=3,S=0,T=5,P=1,bias_strength=0)
        # List of strategies
        STRATEGY_LABELS = (`Cooperate', `Defect') 
        # List of equilibria
        EQUILIBRIA_LABELS=(`Cooperation', `Defection')
    
        def __init__(self,R,S,T,P,bias_strength):
            # Define the payoff matrix
            payoff_matrix = ((R,S),(T,P)) 
            super(PrisonersDilemma, self).__init__(payoff_matrix,1,bias_strength)
        
        @classmethod
        # Function that defines the equilibria
        def classify(cls, params, state, tolerance): 
        
            threshold = 1 - tolerance
            if state[0][0] > threshold:
                return 0 # Cooperate 
            elif state[0][1] > threshold:
                return 1 # Defect
            else:
                return super(PrisonersDilemma, cls).classify(params, state, tolerance)
\end{verbatim}

The Prisoners' Dilemma is a symmetric game, that is, all the players have the same strategies and payoffs. So we use the SymmetricNPlayerGame subclass of the Game class. 

The class method, `classify' is used to identify states of interest, in this case, the state in which all players cooperate, and the state in which all players defect. The classify command identifies whether the population is ``at" these states at the end of a round.  Technically, it tests whether `1 - tolerance' of the population is at the state, where the `tolerance' parameter is chosen by the user (see the SI for a more complicated example that uses the parameters of the game to arrive at the desired tolerances). If the system is in a state other than the ones defined by the user, the classify command assigns the state as `Unclassified'. These states need not be associated with pure strategies only and can include steady states consisting of mixed strategies.

\section*{Types of Simulations}
Once the game is created, the user can choose between either stochastic (Moran, Wright-Fisher) or deterministic (Replicator) dynamics. The GameDynamicsWrapper and VariedGame classes take care of combining the chosen game and dynamics and run the desired simulation for a given number of generations and population size. The user can also specify a `start\_state' which gives the initial strategy frequencies (see SI for details). It defaults to a random list of initial frequencies with the population divided approximately equally amongst all the strategies. \textbf{Table \ref{tab:classes}} describes some of the important classes in DyPy. We provide some commonly used simulation methods in the package which are enumerated below along with examples.

\begin{table}[t]
    \centering
    \begin{tabular}{m{4cm}|m{7cm}}
        \hline
        Class & Description \\
        \hline
        Game & Encapsulates the idea of the game that is to be simulated. The user can create a subclass of this class and define the game by specifying the payoff matrix, number of players and strategies along with the equilibria of interest. \\ \hline
        Dynamics & Dynamics govern the update rules from one generation to the next. We provide three commonly used dynamics: Moran, Wright-Fisher and Replicator. The user can also define their own. \\ \hline
        GameDynamicsWrapper & A helper class that wraps a dynamics class and a game class. It provides helper methods for simulation with a fixed set of game and dynamics parameters.  \\ \hline
        VariedGame & A helper class that wraps a dynamics class and a game class. It provides helper methods for simulation while varying one or more parameters.
    \end{tabular}
    \caption{Brief description of important DyPy classes}
    \label{tab:classes}
\end{table}

\subsection*{A single simulation of the game's dynamic}

\begin{figure}[htbp]
    \centering
    \includegraphics[scale=1.0]{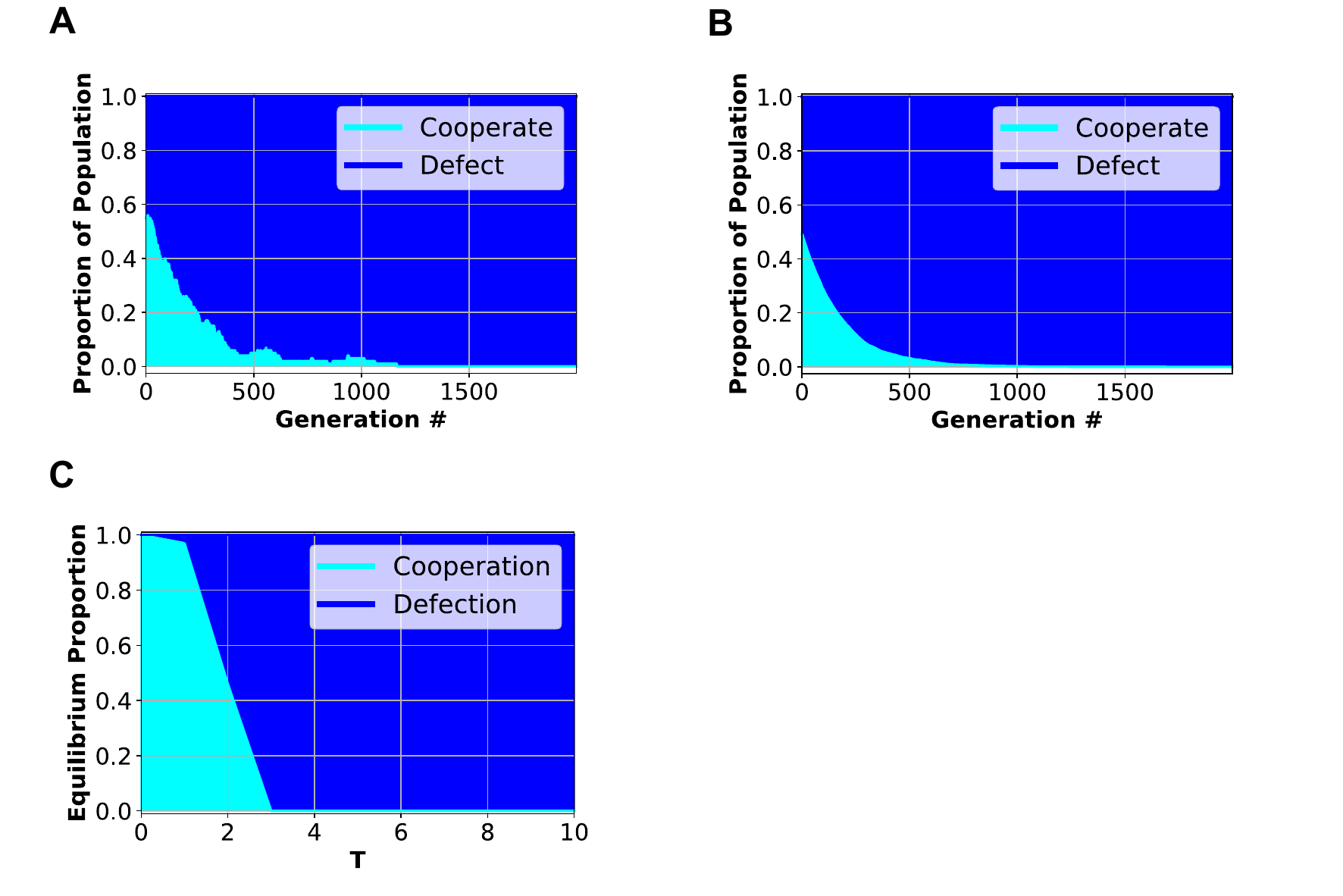}
    \caption{\textbf{Simulation results for a population playing the Prisoners' Dilemma using the three main methods in DyPy.} A) Evolution of strategies via the Moran process for 2000 generations. B) Evolution of strategies via the Moran process for 2000 generations averaged over 100 iterations. C) The equilibrium proportion of strategies on varying the payoff $T$ in the Prisoners' Dilemma under Moran dynamics. Each simulation was run for 4000 generations averaged over 100 iterations per value of $T$ $(1,2,\dots, 10)$. We used a population size of 100 for all the simulations with the payoffs in the Prisoners' Dilemma set to $R=3$, $S=0$, $T=5$ and $P=1$.}
    \label{fig:simplePD}
\end{figure}

The `simulate' method can be used when the user wishes to analyze the dynamics of strategies in a single EGT simulation for a specified number of generations (number of updating steps in the entire population). In both stochastic dynamics, the user can also specify a mutation rate, $\mu$ (see SI for more details). The result is a graph of the dynamics of each players' strategies over time. For example, we simulate the Prisoners' Dilemma game using Moran dynamics in the absence of mutations. The output is a graph, \textbf{Figure \ref{fig:simplePD}A}. We see that the dynamics converge to defection which is the Nash equilibrium. Example code:

\begin{verbatim}
from wrapper import GameDynamicsWrapper
from dynamics.moran import Moran
s = GameDynamicsWrapper(PrisonersDilemma,Moran,dynamics_kwargs={`mu':0})
s.simulate(num_gens=2000,pop_size=100,graph=dict(area=True,options=[`smallfont']))
\end{verbatim}

We provide an example of using the deterministic Replicator dynamic in the Supplementary Information.

\subsection*{Multiple iterations of a simulation of the game's dynamic}

On many occasions we want to run a simulation multiple times in order to check the robustness of the results in the presence of variations in the initial conditions and/or due to the inherent stochasticity of the dynamics. The game may also have multiple equilibria, in which case the user might be interested in the fraction of generations that the population spends in each. For these reasons, we provide the method `simulate\_many' where multiple iterations of the simulation are run and the frequency of each resulting equilibrium is returned. If the system is not within `tolerance' of a state defined using the classify command, the simulation returns `Unclassified'. When multiple simulations are performed, they are automatically parallelized across all available cores. 

We now use `simulate\_many', and the stochastic Moran dynamic, to show that a population playing the Prisoners' Dilemma stabilizes at defection even in the presence of stochasticity.

\begin{verbatim}
from wrapper import GameDynamicsWrapper
from dynamics.moran import Moran
s = GameDynamicsWrapper(PrisonersDilemma,Moran)
s.simulate_many(num_iterations=100, num_gens=2000,pop_size=100,\
graph=dict(area=True, options=[`smallfont']))
\end{verbatim}

This command returns a text output with the frequency of each equilibrium,

\begin{verbatim}
{`Defection':1.0}
\end{verbatim}

The simulation never returns `Unclassified' and defection equilibrium is reached reliably during each iteration. It also returns a graph (\textbf{Figure \ref{fig:simplePD}B}) which is an average over the iterations of the dynamics of each players' strategies over time. In the SI we provide an example where `simulate\_many' is used for a game (Repeated Prisoners' Dilemma) that exhibits multiple stable equilibria.

\subsection*{Approximating the fixation probability}

The `frac\_invasions' method computes the fraction of iterations where a strategy introduced in a population, dominates other strategies, after a specified number of generations. The user can define the amount of prevalence needed for a strategy to be considered `dominating' via the `tolerance' parameter.

We include this method because it can be used to approximate the fixation probability for a strategy. In a game with two strategies A and B, the fixation probability of A is the probability that it fixes in the population when introduced in a population of only B players. This concept is used extensively in the field of population genetics, for example, to study how a particular allele may fix in the population \citep{wahl2008, DEOLIVEIRA2004, LAMBERT2006}.

The fixation probability for a strategy can be approximated by setting the tolerance $=0$ and running the simulation for an appropriate number of generations. This is to ensure that the simulation is run long enough such that the strategy has either fixed in the population or the population has reached some other equilibrium. A thorough investigation of the analytical, numerical and simulation methods to compute the fixation probability is given in \cite{Hindersin2019}. The techniques described can be used to decide the length of the simulation. The following few lines of code can be used to approximate the fixation probability for `Defect' in the Prisoners' Dilemma with tolerance set to $0$,

\begin{verbatim}
from wrapper import GameDynamicsWrapper
from dynamics.moran import Moran
s = GameDynamicsWrapper(PrisonersDilemma,Moran)
s.frac_invasions(num_iterations = 1000, num_gens = 2000, \ 
pop_size = 100, strategy_indx = 1) # Index of strategy `Defect'
\end{verbatim}

This returns a text output of the form :

\begin{verbatim}
`Fraction of runs where the required strategy dominated the population = 0.79'
\end{verbatim}

The fixation probabilities of the strategies in the Prisoners' Dilemma can be analytically computed (\cite{Hindersin2019}). For the payoffs and selection strength used in our simulation, the fixation probability for `Defect' is 0.8 which is in close agreement with the approximation obtained via the simulation.

\subsection*{Effect of varying parameters}

We provide the `vary' method to analyze the effect of varying one or more parameters associated with the dynamics or with the game. For each value of the parameter(s) being varied, multiple iterations of the simulation are run and the final frequency of each resulting equilibrium is recorded. The output is a graph of these final frequencies as a function of the varied parameter(s). As an example we vary the payoff value $T$ $(1,2,\dots, 10)$ from the Prisoners' Dilemma Game. 

\begin{verbatim}
from wrapper import VariedGame
from dynamics.moran import Moran
s = VariedGame(PrisonersDilemma, Moran)
s.vary(game_kwargs={`T':[0,10,10]},num_gens=4000,\ 
num_iterations=100,graph=dict(area=True,options=[`smallfont']))
\end{verbatim}

The output, \textbf{Figure \ref{fig:simplePD}C} shows the equilibrium proportion of strategies on varying $T$ with fixed $R=3$. The game is a Prisoners' Dilemma when $T>R$ and hence, we see the dynamics converging to defection once this is true.

In addition to EGT simulations in well-mixed populations with payoff-based learning we also include the possibility to simulate effects of frequency biased imitation and group selection.

\section*{Frequency Biased Imitation}

One of the motivations for using evolutionary dynamics is that humans learn or imitate, and that, crucially, they preferentially learn or imitate successful strategies  \citep{chudek2012prestige,harris2011young,stenberg2009selectivity,galef2008useof,laland2011experimental}, just as in biological evolution, natural selection favors successful strategies.  Thus, the dynamics we've focused on so far (Replicator, Wright-Fisher, and Moran) can be used to describe learning and imitation in the same way they were used to describe biological evolution.

However, humans are sometimes known to imitate \textit{common} strategies, somewhat independently of whether they are successful \citep{boyd1985culture,chudekcultural}.  This is known as \textit{frequency biased imitation}.  If the frequency bias is important, relative to the success bias, this can change how a dynamic will behave, and where it will stabilize.  

For these reasons, DyPy makes it possible to include frequency biased imitation when analyzing the dynamics of a game.  We incorporate frequency biased imitation by adding to the average payoff from playing a strategy, a user-specified function of the strategy's frequency: $u_i^{FBI}(\sigma,\sigma^{\prime}_{1},\dots ,\sigma^{\prime}_{N}) = u_i(\sigma, \sigma^{\prime}_{1},\dots, \sigma^{\prime}_{N})(1-\gamma) + \phi(\sigma)\gamma$ where $u_i(\sigma, \sigma^{\prime}_{1}, \dots,\sigma^{\prime}_{N})$ is player $i$'s average payoff from playing $\sigma$ in a population when playing opponents with the strategy set $\left\{\sigma^{\prime}_{1}, \dots, \sigma^{\prime}_{N}\right\}$, $\gamma$ is the relative strength of individual versus frequency biased (conformist) imitation and $\phi(\sigma)$ is the user-specified function of $\sigma$'s frequency.  This function defaults to the commonly used function \citep{nakahashi2007},

\begin{equation}
\phi(\sigma)=\left(\frac{x_{\sigma}^{a}}{\sum_{j=1}^{M}x_{j}^{a}}\right)\times s 
\end{equation}
where $x_{\sigma}$ is the frequency of strategy $\sigma$, the sum is over the frequencies of all $M$ strategies that player $i$ can employ, $s$ is a scaling factor and $a$ is the `conformist' parameter as defined by Nakahashi \citep{nakahashi2007}. $a=1$ is the default in the package. In the code we refer to s as the `bias\_scale' and $\gamma$ as the `bias\_strength'. This formalism is a more general version of that used in \citep{molleman2013}, where the effect of conformity is thought of as a coordination game.\newline
\newline \textit{Conformism in Prisoners' Dilemma}
\newline

Conformism is a social learning strategy where players learn strategies by imitating the majority \citep{boyd1985culture}. Adding the effects of conformism to payoff-based learning considered so far can help stabilize cooperation under certain conditions. We can simulate this system for a Prisoners' Dilemma Game via the VariedGame class and vary the strength of conformism to reproduce the results from \citep{molleman2013}. See SI for all the parameters used in the simulation.

\begin{verbatim}
from wrapper import VariedGame
from dynamics.wright_fisher import WrightFisher
s = VariedGame(PrisonersDilemma,WrightFisher)
s.vary(game_kwargs={`bias_strength':[0,1,10]},num_gens=100,num_iterations=100,\ 
parallelize=True,graph=dict(area=True,options=[`smallfont']))
\end{verbatim}

\begin{figure}[htbp]
    \centering
    \includegraphics[scale=0.95]{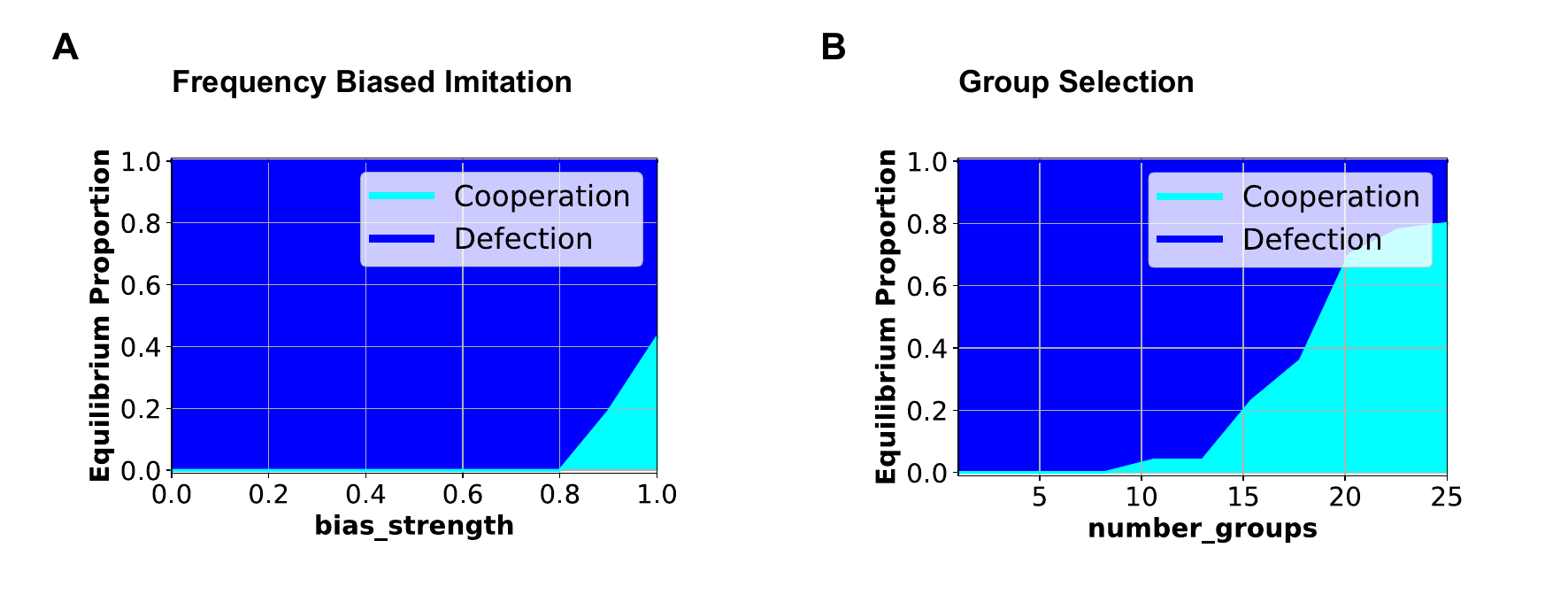}
    \caption{\textbf{Equilibrium proportion of strategies in the presence of frequency biased imitation and group selection.} \textbf{Left:} The proportion of times an equilibrium arises in the population while varying the strength of conformism in a Prisoners' Dilemma game under Wright-Fisher dynamics. $\gamma$ (bias\_strength) is varied from 0 (no conformism) to 1 (fully conformist learning). Each simulation was run for 100 generations, averaged over 100 iterations per value of $\gamma$. \textbf{Right:} The proportion of times an equilibrium arises in the population playing the Prisoners' Dilemma (Wright-Fisher dynamics) on varying the number of groups and group size keeping the total population fixed. The number of groups was varied from 1 to 25. At each group number, the simulation was run for 100 generations, each averaged over a 100 iterations. The `rate' of group selection was 0.2. The population size was set to 100 for both simulations.}
    \label{fig:2}
\end{figure}

In \textbf{Figure \ref{fig:2}A}, as the strength of conformism increases, cooperation arises in the population. When the learning is entirely conformist ($\gamma = 1$), the population converges to cooperation or defection solely depending upon the initial state and so we see both strategies fixing $\sim 50\%$ of the time.

\section*{Group Selection}
Group selection is the idea that sometimes evolution occurs via natural selection acting at the level of a group in addition to the individual level \citep{LUO201441} which leads to favoring traits (strategies) that are advantageous to the group as a whole. In the presence of such multilevel (group and individual) selection, a games' dynamics may no longer converge to the Nash equilibrium. For example, \citep{Traulsen10952} showed that cooperation is favored over defection under certain conditions in the presence of group selection. With this in mind, we include tools to simulate such multilevel selection dynamics in DyPy.

We use the formalism of group selection \citep{LUO201441} that consists of a fixed number of groups $m$ each consisting of a fixed population of size $n$. At each time-step, the system can undergo either group or individual-level selection via a Moran or Wright-Fisher process. The antagonism between the two levels of selection is incorporated via a `rate' which is the probability that group selection occurs at each time-step. At the individual-level, reproduction occurs proportional to the individual fitness whereas, groups reproduce proportional to their average fitness. Specifically, in the Moran process during each round of replication, one individual (group) from the entire population is chosen for reproduction proportional to their fitness. On the other hand, all individuals (groups) reproduce proportional to their fitness during each round of replication in the Wright-Fisher process. The population (group) number is kept fixed by randomly selecting individuals (groups) to die whenever reproduction occurs. \newline
\newline \textit{Group Selection in Prisoners' Dilemma}
\newline

Group selection is another mechanism by which cooperation might be stabilized in the population. Cooperation can be favored in a population with multi-level selection \citep{LUO201441,Traulsen10952} where competition between groups leads to cooperative behaviour under certain conditions. It has been observed that smaller group sizes and large number of groups favor cooperators \citep{Traulsen10952,Traulsen2008}. We can simulate this in our package by using the VariedGame class and the Prisoners' Dilemma game. This simulation varies the number of groups in the population keeping the total population size fixed. We simulate it over multiple iterations for a fixed group number. 

\begin{verbatim}
from wrapper import VariedGame
from dynamics.wright_fisher import WrightFisher
s = VariedGame(PrisonersDilemma, WrightFisher,dynamics_kwargs={`rate':0.2})
s.vary(dynamics_kwargs={`number_groups':[1,25,10]},num_gens=100,\ 
num_iterations=100,graph=dict(area=True,options=[`smallfont']))
\end{verbatim}

The output is a graph (\textbf{Figure \ref{fig:2}B}) showing the proportion of times a steady state is attained in the simulation. As the number of groups increases and the group sizes decrease, we see cooperation arising in the population. 

\section*{Rock-Paper-Scissors}

\begin{figure}
    \centering
    \includegraphics[scale=0.95]{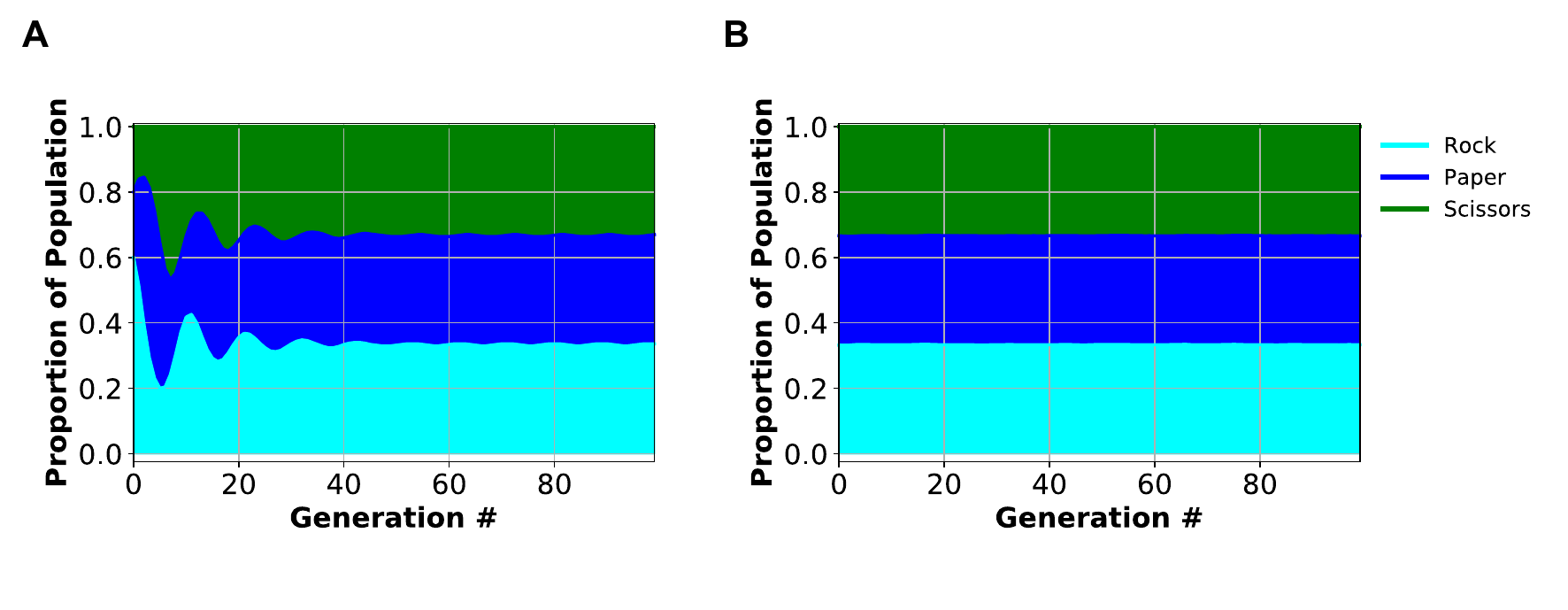}
    \caption{\textbf{Evolution of strategies for a population (n=500) playing the Rock-Paper-Scissors game using the Wright-Fisher dynamic.} (A) Results of a single run of the simulation for 100 generations. (b) An average over 500 iterations of the simulation in the presence of a $3\%$ mutation rate.  `simulate\_many' also returns the text output,
    \{`Nash': 1.0\} (see SI for how this is defined), implying that the Nash equilibrium was reached at the end of each iteration.}
    \label{fig:RPS}
\end{figure}

So far we have analyzed dynamics in the Prisoners' Dilemma where, in equilibrium, all players play $D$. We now present, in brief, the dynamics of the Rock-Paper-Scissors (RPS) game, for which the only equilibrium is mixed with one-third of players to play rock, one-third to play paper, and one-third to play scissors.  An analysis of the game's dynamics reveals something interesting that the equilibrium analysis could not: the population often cycles for some time, or even forever, depending on the relative size of the payoffs for winning vs. losing a round of the game \citep{hoffman2015experimental}. We illustrate this in \textbf{Figure \ref{fig:RPS}} using a stochastic process including mutations, and the `simulate' and `simulate\_many' methods (see SI for the code and parameters used).

\section*{Discussion}
This paper presents DyPy, a Python library for facilitating a variety of useful EGT simulations, alongside a few simple examples to illustrate the library's functionality.  
We hope DyPy will prove to be an asset for experienced and novice researchers alike, as well as for teachers.  
We plan to continue to maintain and expand DyPy's functionality, for instance, by adding additional dynamics such as reinforcement learning and \citet{fudenberg2006imitation}'s imitation process with rare mutations, as well as permitting structured populations.  We encourage users to contact us with suggestions for additional functionality, and also encourage pull requests on GitHub so that users can help with the development of the library.

\section*{Author Contributions}
DyPy was conceived by EY and EL. It was originally written by EL in Python 2 which was updated to Python 3 along with additional functionality by AF. AN extended the package to include frequency biased imitation, group selection and additional functionalities. AN led the writing of the manuscript under the guidance of EY and MN. All authors contributed to the draft and gave their approval for publication. 
\section*{Acknowledgments}
We would like to thank Alexander Heyde, Christian Hilbe and Laura Schmid for helpful discussions and edits regarding the manuscript. This work was funded partly by NIH DP5OD019851 (Anjalika Nande).
\bibliographystyle{plainnat}
\bibliography{DyPy.bib}
\end{document}


\maketitle
\section{Introduction}

This SI includes additional useful information on how to use the package and provides example code for a slightly more complicated game - Repeated Prisoners' dilemma - than the Prisoners' Dilemma in the main text. In addition, we provide the code that was used in the conformity and Rock-Paper-Scissors examples in the main text. We also focus on how to define a start\_state, incorporate the effect of mutations and list additional graphing options that are provided in the package.

\section{Defining the start\_state}

The `start\_state' corresponds to the initial frequencies of strategies that can be specified by the user. It is a multidimensional list of size $m \times n \times l$ where $m$ is the number of groups, $n$ is the number of player types and $l$ is the number of player strategies. For example, in the Prisoners' Dilemma game defined in the main text, the start\_state would be of the following form in the absence of group selection,

\begin{verbatim}
start_state = [[[60,40]]]
\end{verbatim}

\noindent This means that in a total population of $100$ individuals, at the start there are $60$ individuals playing `Cooperation' whereas $40$ individuals are playing `Defection'. The order of the two strategies is dependent upon how it was defined in the game. The start state can be included when using any of the simulation methods, for example,

\begin{verbatim}
s.simulate(num_gens=100,pop_size=100,start_state=[[[60,40]]])
\end{verbatim}

\section{Mutations}

We provide functionality to include the effects of mutations in the two stochastic dynamics - Wright-Fisher and Moran. The user can either provide a universal mutation rate which applies to each strategy or a list of mutation rates with each strategy of each player getting its own mutation rate. For example, in a 2 player game with Player 1 having 2 strategies and Player 2 having 3 strategies, the mutation matrix takes the form,

\begin{verbatim}
mu = [[0.1,0.2],[0.1,0.2,0.3]]
\end{verbatim}

\noindent and can be specified while initializing the GameDynamicWrapper or VariedGame classes, for example,

\begin{verbatim}
s = GameDynamicsWrapper(PrisonersDilemma,Moran,\ 
dynamics_kwargs={`mu':[[0.1,0.2],[0.1,0.2,0.3]]})
\end{verbatim}
\section{Repeated Prisoners' Dilemma}

\begin{figure}[htbp]
    \centering
    \includegraphics[scale=0.9]{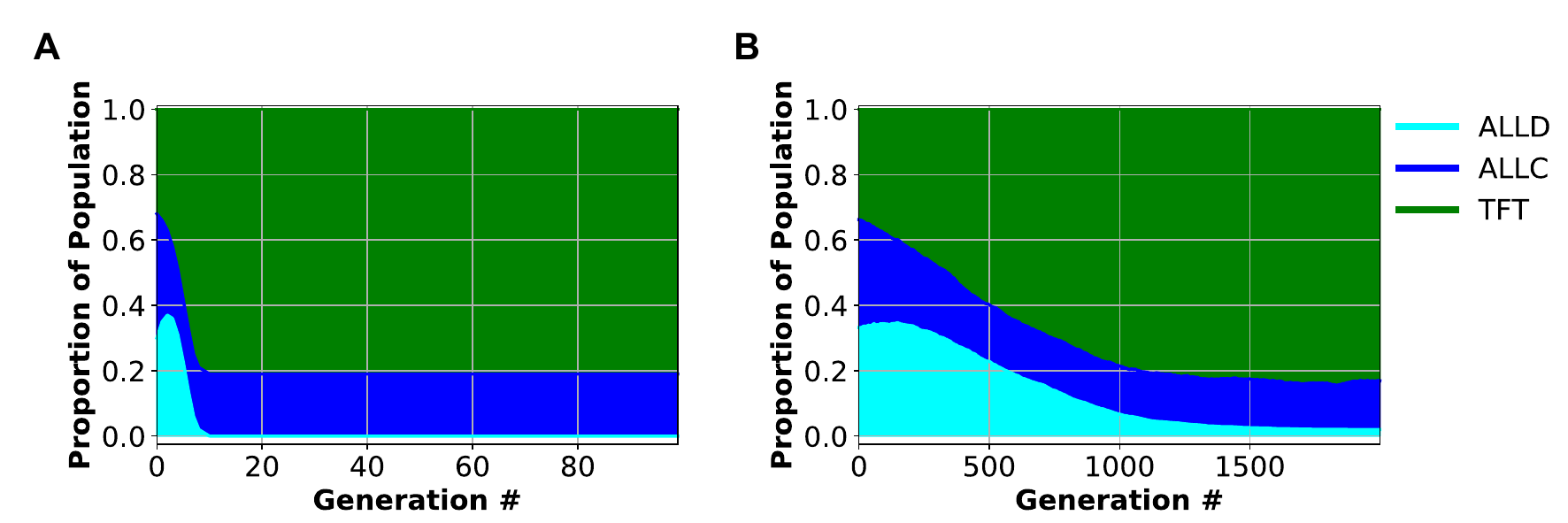}
    \caption{\textbf{Evolution of strategies for a population playing repeated Prisoners' Dilemma strategies: ALLD, ALLC and TFT.} A) Under Replicator dynamic for 100 generations. B) Under the Moran dynamic for 2000 generations, averaged over 100 iterations. The population size was set to 100 for both the simulations.}
    \label{fig:repeatedPD}
\end{figure}

Different approaches have been utilized to stabilize cooperation in the standard Prisoners' Dilemma \citep{nowak2006evolutionary}. `Repeated' games is the idea that the game is not played just once but is repeated several times between two players. For such games, there are strategies such as the Tit-for-tat (TFT) strategy that are stable against invasion by Always defect (ALLD) \citep{axelrod1980,axelrod21980}. TFT starts with cooperation and then plays for the subsequent rounds whatever strategy the opponent played in the previous round. We can simulate a population playing TFT, ALLD and Always Cooperate (ALLC) and show that TFT helps maintain cooperation in the absence of mistakes. In this section we illustrate the results of the simulations. See the next section for how the game (RepeatedPD) is coded along with how we defined the expected equilibrium states. We first simulate the population using the Replicator dynamic,

\begin{verbatim}
s = GameDynamicsWrapper(RepeatedPD,Replicator) 
s.simulate(num_gens = 100, pop_size = 100, graph=dict(area=True,options=[`smallfont']))
\end{verbatim}

The result of this simulation is \textbf{Figure \ref{fig:repeatedPD}A} which shows that TFT helps ALLC dominate ALLD. Next, we check the robustness of this final state in the presence of stochasticity.

\begin{verbatim}
s = GameDynamicsWrapper(RepeatedPD,Moran) 
s.simulate_many(num_iterations = 100 , num_gens = 2000, pop_size = 100,\
class_end = True, graph=dict(area=True,options=[`smallfont']))
\end{verbatim}

This command returns a text output with the proportion of each equilibrium.

\begin{verbatim}
{`Cooperative Equilibrium': 0.98,`Non Cooperative Equilibirum': 0.02}
\end{verbatim}

The plot in \textbf{Figure \ref{fig:repeatedPD}B} is the average over all iterations for each generation.

\section{Code for the Repeated Prisoners' Dilemma}

Always Defect (ALLD), Always Cooperate (ALLC) and Tit-for-tat (TFT) belong to the set of reactive strategies and the entries in the payoff matrix for these three strategies playing each other, Eq.(\ref{eq:RepeatedPD}) can be calculated via well established methods \citep{nowak2006evolutionary}. 

\begin{equation} \label{eq:RepeatedPD}
\bordermatrix{~ & ALLD & ALLC & TFT \cr
  ALLD & P & T & P \cr
  ALLC & S & R & R \cr
  TFT  & P & R & R \cr}
\end{equation}

Here $T>R>P>S$ are the standard payoffs associated with the basic Prisoners' Dilemma game from Eq.(2) in the main text.
We define the `Cooperative' equilibrium as the state of the system where ALLD can't invade TFT. ALLD can invade TFT if the fitness of an ALLD player is higher than that of TFT. In Evolutionary Game Theory, fitness is defined to be proportional to the payoffs. So if $x_{ALLD}$, $x_{ALLC}$ and $x_{TFT}$ are the frequencies of ALLD, ALLC and TFT players respectively, the fitness of each player is given by,

\begin{align} \label{eq:RepeatedPDfit}
    f_{ALLD} &\propto  P x_{ALLD} + T x_{ALLC} + P x_{TFT}  \\
    f_{ALLC} &\propto S x_{ALLD} + R x_{ALLC} + R x_{TFT}  \\
    f_{TFT} &\propto P x_{ALLD} + R x_{ALLC} + R x_{TFT}
\end{align}

 The state of the system can be `Cooperative' as long as $f_{ALLD}<f_{TFT}$. Solving this equation we find that the system is in a cooperative equilibrium when,

\begin{equation}
    x_{ALLC} < \frac{(R-P)}{(T-R)} x_{TFT}
\end{equation}

 When this condition is not met, ALLD can invade TFT and this leads to a `Non-Cooperative' equilibrium that is dominated by defectors. Once the payoff matrix and the equilibria are defined, we can code the game as follows. 

\begin{verbatim}
from games.game import SymmetricNPlayerGame
class RepeatedPD(SymmetricNPlayerGame):

    DEFAULT_PARAMS = dict(R=3, T=5, S=0, P=1, bias_strength=0)
    STRATEGY_LABELS = (`ALLD', `ALLC', `TFT')
    EQUILIBRIA_LABELS = (`Cooperative Equilibrium', `Non Cooperative Equilibirum')

    def __init__(self, R, T, S, P, bias_strength):

        
        payoff_matrix = ((P, T, P),
                         (S, R, R),
                         (P, R, R))
        

        super(RepeatedPD, self).__init__(payoff_matrix, 1, bias_strength)
        
    
    @classmethod
    def classify(cls, params, state, tolerance):
    
        R = getattr(params,"R")
        T = getattr(params,"T")
        S = getattr(params,"S")
        P = getattr(params,"P")
        
        tolerance = (R-P)/(T-R)
        if state[0][1] <= tolerance * state[0][2] :
            return 0 # Cooperative Equilibrium
        elif state[0][1] > tolerance * state[0][2] :
            return 1 # NonCooperative Equilibrium
        else:
            return super(RepeatedPD, cls).classify(params, state, tolerance)
\end{verbatim}

\section{Code for the Conformity Example}

In order to compare our results qualitatively with \citep{molleman2013}, we chose the `bias\_function' to be linear in frequency, $\phi(x)=2x$ (`bias\_scale' is set to 2) such that conformity is modelled as a coordination game.\footnote{`bias\_function' is defined in the `get\_expected\_payoff' method of the `PayoffMatrix' class.} For this modified Prisoners' Dilemma, the fitness of each player which is proportional to their payoffs is given by,

\begin{align}\label{eq:PDconformity}
f_C &\propto (Rx_C + Sx_D)(1-\gamma)+2 \gamma x_C \\
f_D &\propto (Tx_C+Px_D)(1-\gamma)+2 \gamma x_D
\end{align}

with $T>R>P>S$ as the usual payoffs associated with the Prisoners' Dilemma. As in the previous example, we define the `Cooperative' equilibrium of the system to be when defectors can't invade cooperators that is, $f_C>f_D$. Solving Eq.(\ref{eq:PDconformity}) this leads to the following conditions for a cooperative equilibrium on the frequencies of the two strategies,

\begin{equation}\label{eq:conformityeq}
x_C > x_D \frac{(P-S)(1-\gamma)+2\gamma}{(R-T)(1-\gamma)+2\gamma}
\end{equation}

with the caveat that such an $x_C$ exists only when $(R-T)(1-\gamma)+2\gamma>0$ since $x_C,x_D > 0$.
Exact code used to generate the game:

\begin{verbatim}
from games.game import SymmetricNPlayerGame
class PrisonersDilemma(SymmetricNPlayerGame):
    DEFAULT_PARAMS = dict(R=3, S=0, T=5, P=1, bias_strength=0.0, bias_scale=2)
    STRATEGY_LABELS = (`Cooperate', `Defect')
    EQUILIBRIA_LABELS = (`Cooperation',`Defection')
    def __init__(self,R,S,T,P,bias_strength,bias_scale):
        payoff_matrix = ((R,S),(T,P))
        super(PrisonersDilemma, self).__init__(payoff_matrix, 1, bias_strength,\
        bias_scale)
        
    @classmethod
    def classify(cls, params, state, tolerance):
        
        R = getattr(params,"R")
        T = getattr(params,"T")
        S = getattr(params,"S")
        P = getattr(params,"P")
        bias = getattr(params,"bias_strength")
        
        # Avoid division by zero
        if bias !=0.5:
            ratio = ((P-S)*(1-bias) + 2*bias)/((R-T)*(1-bias) + 2*bias)
        else:
            ratio = 1
            
        if ratio > 0:
            if state[0][0] > state[0][1]*ratio:
                return 0 # Cooperation
            elif state[0][0] <= state[0][1]*ratio:
                return 1 # Defection
            else:
                return super(PrisonersDilemma, cls).classify(params, state, tolerance)
        if ratio < 0:
            return 1     
\end{verbatim}

\section{Rock-Paper-Scissors}

Rock-Paper-Scissors (RPS) is a three strategy game where there is a cyclic domination of strategies. Rock beats scissors which beats paper which in turn beats rock. Any symmetric three strategy game with cyclic domination characterizes a RPS game (\cite{nowak2006evolutionary}). We use the following payoff matrix in our simulations,

\begin{equation} \label{eq:RSP}
\bordermatrix{~ & R & P & S \cr
  R & 0 & -0.2 & 2 \cr
  P & 2 & 0 & -0.2 \cr
  S  & -0.2 & 2 & 0 \cr}
\end{equation}

The Nash equilibrium for this game is a mixed strategy with a uniform distribution over the pure strategies, $(\frac{1}{3},\frac{1}{3},\frac{1}{3})$. The code used to generate the game,

\begin{verbatim}
class RPS(SymmetricNPlayerGame):
    DEFAULT_PARAMS = dict(a1=0.2, a2=0.2, a3=0.2, b1=2, b2=2, b3=2)
    STRATEGY_LABELS = (`Rock', `Paper', `Scissors')
    EQUILIBRIA_LABELS = (`Nash',)
    def __init__(self, a1, a2, a3, b1, b2, b3):
        payoff_matrix = ((0, -a2, b3),(b1, 0, -a3),(-a1, b2, 0))
        super(RPS, self).__init__(payoff_matrix,1)
        
    @classmethod
    def classify(cls, params, state, tolerance):
        
        threshold = 1/3
        pop_rock = state[0][0]
        pop_paper = state[0][1]
        pop_scissors = state[0][2]
        
        if abs(pop_rock-threshold) <= tolerance and \ 
        abs(pop_paper-threshold) <= tolerance and \
        abs(pop_scissors-threshold) <= tolerance:
            return 0 #Nash
        else:
            return super(RSP, cls).classify(params, state, tolerance)
\end{verbatim}

If the population converges to the Nash equilibrium, the `classify' method returns `Nash', otherwise it returns `Unclassified'. We set the tolerance to $0.03$. The code used to generate the single iteration of the simulation,

\begin{verbatim}
s = GameDynamicsWrapper(RSP,WrightFisher)
s.simulate(num_gens=100, pop_size=500, start_state = [[[300,100,100]]],\
graph = dict(area=True,options=[`smallfont']))
\end{verbatim}

We use this start state to show that the fixed point is stable even if the dynamics start further from the equilibrium point.  The code used to generate the multiple iterations of the simulation in the presence of mutations,

\begin{verbatim}
s = GameDynamicsWrapper(RSP, WrightFisher, dynamics_kwargs={`mu':0.03})
s.simulate_many(num_iterations = 500, num_gens = 100, pop_size=500, \ 
graph = dict(area=True,options=[`smallfont']))
\end{verbatim}

\section{Other Graphing Options}
The library provides other graphing methods which were not included in the main text. We use the basic Prisoners' Dilemma game to showcase how these methods may be used.

\subsection{Histogram}

\begin{figure}[htbp]
    \centering
    \includegraphics[scale=0.4]{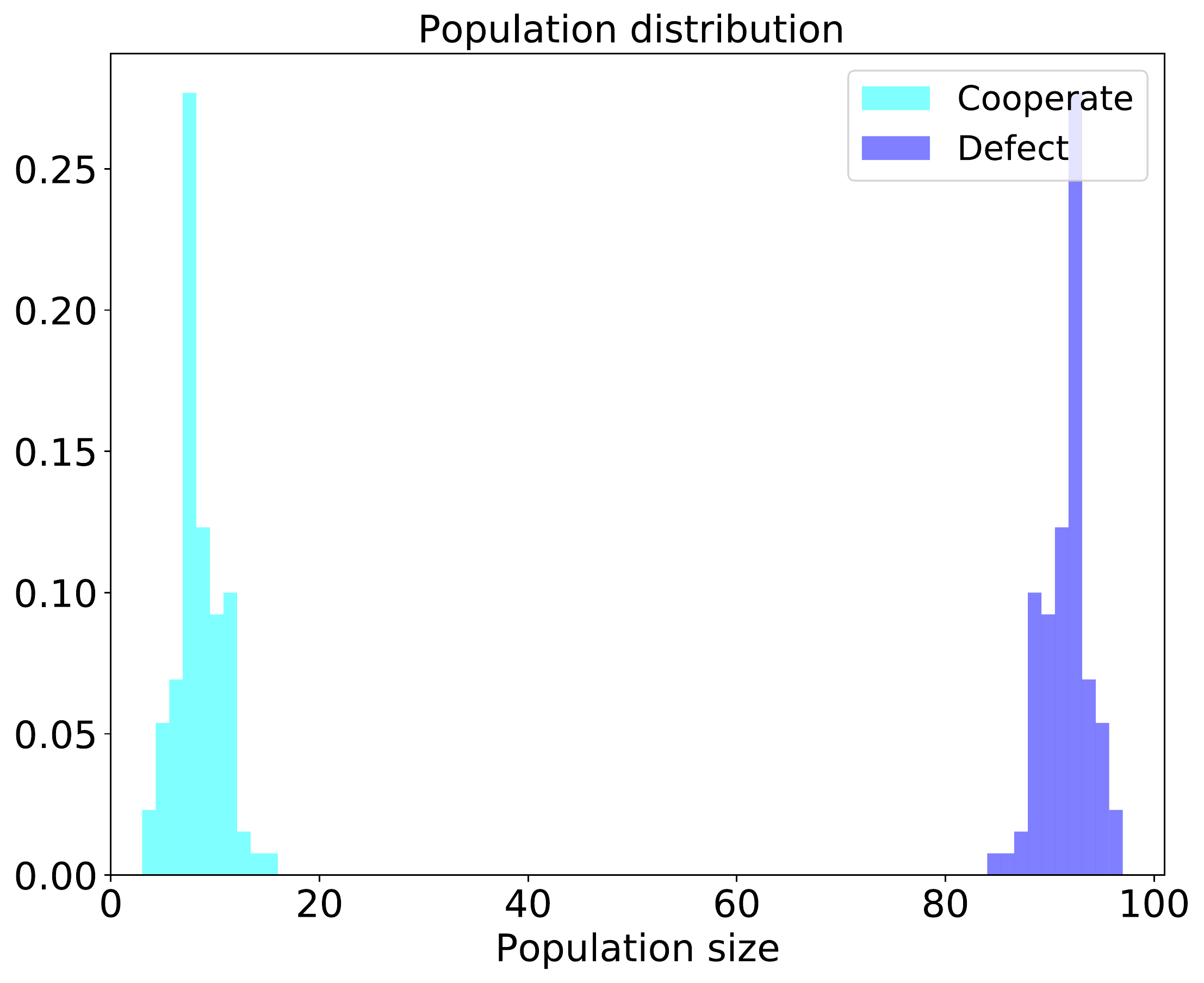}
    \caption{\textbf{Distribution of the final population sizes for the strategies in the Prisoners' Dilemma under Wright-Fisher dynamics.} The results are for 100 iterations each of 100 generations, for a populations size of 100. This was in the presence of a $10\%$ mutation rate for both the strategies.}
    \label{fig:histogramPD}
\end{figure}

We provide the option to plot histograms of the distribution of the strategies for each player at the end of the iterations in the method `simulate\_many'. For example, the code below gives the output in \textbf{Figure \ref{fig:histogramPD}}.

\begin{verbatim}
s = GameDynamicsWrapper(PrisonersDilemma, WrightFisher,dynamics_kwargs={`mu':[0.1,0.1]})
s.simulate_many(num_iterations = 100, num_gens = 100, histogram = True)
\end{verbatim}

\subsection{Contour Plots}

\begin{figure}[htbp]
    \centering
    \includegraphics[scale=0.4]{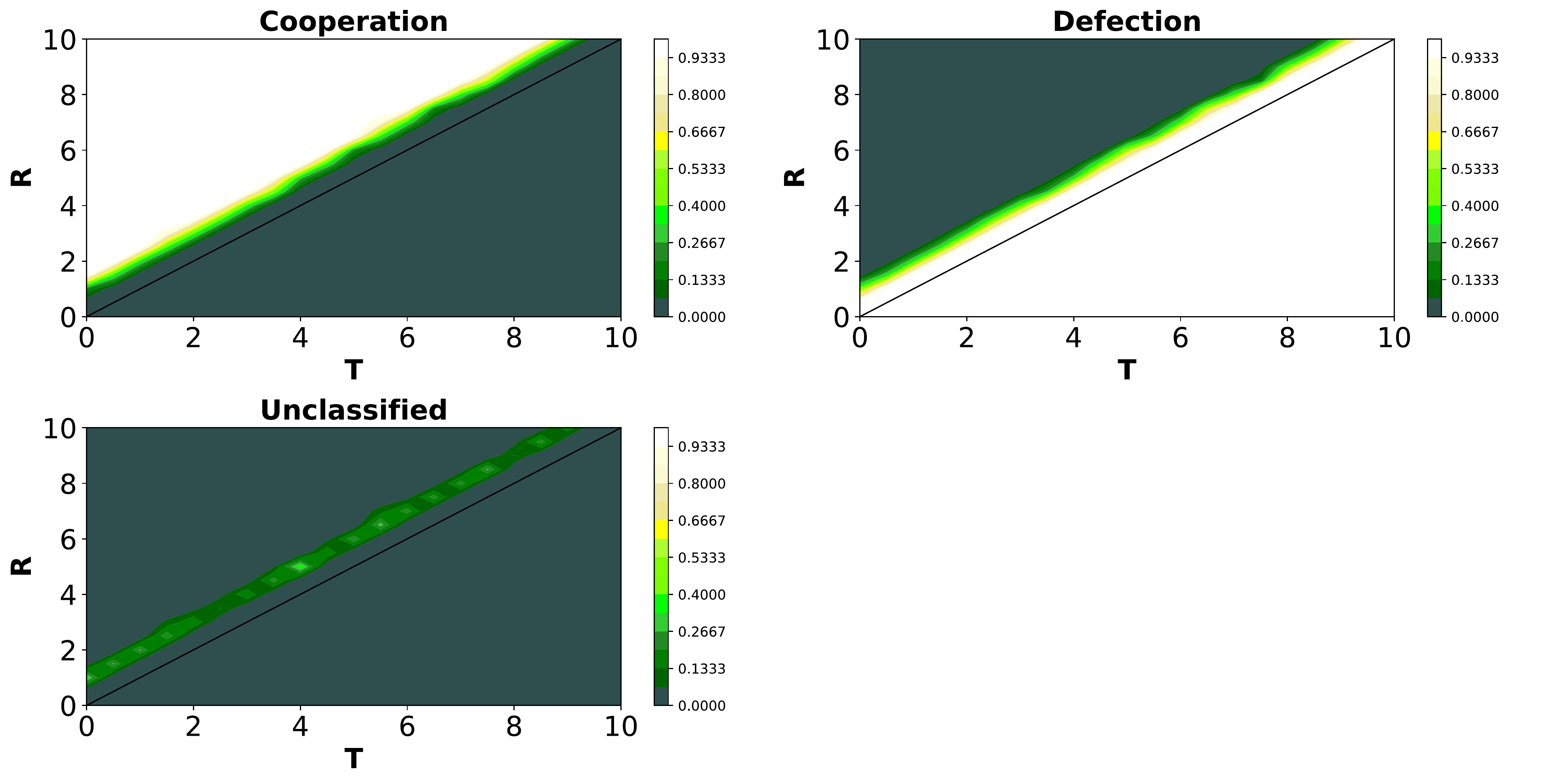}
    \caption{\textbf{Contour plots of the distribution of equilibria on varying the parameters $R$  and $T$ using Wright-Fisher dynamics.} The game is a Prisoners' Dilemma when $T>R$ with `Defection' as the only Nash equilibrium. `Unclassified' corresponds to a mixed equilibrium where both strategies co-exist in the population. Simulations were run for 100 generations with results for each pair of $R$ $[0,10]$ and $T$ $[0,10]$ values averaged over 50 iterations. The solid black line corresponds to $R=T$.}
    \label{fig:contour}
\end{figure}

We can create a contour plot by varying two payoffs, say R and T from the Prisoners' Dilemma. We use the `vary\_2params' helper function to vary two parameters of the game instance. Defection always dominates cooperation as long as $T>R$ (by definition of a Prisoners' Dilemma) so we include the line $T=R$ in the contour plots (\textbf{Figure \ref{fig:contour}}) for reference.

\begin{verbatim}
s = VariedGame(PrisonersDilemma, WrightFisher)
s.vary_2params(`R', (0, 10, 20), `T', (0, 10, 20), num_iterations=50,\
num_gens=100, graph=dict(type=`contour', lineArray=[(0, 10, 0, 10)]))
\end{verbatim}

The game is a Prisoners' Dilemma when $T>R$ and `Defection' is the only Nash equilibrium. When $R>T$ since $S<P$, the system is bistable \citep{nowak2006evolutionary}. In this case, depending upon the initial conditions the system will either converge to cooperation or defection.

\subsection{3D Wire Plots}

\begin{figure}[htbp]
    \centering
    \includegraphics[scale=0.7]{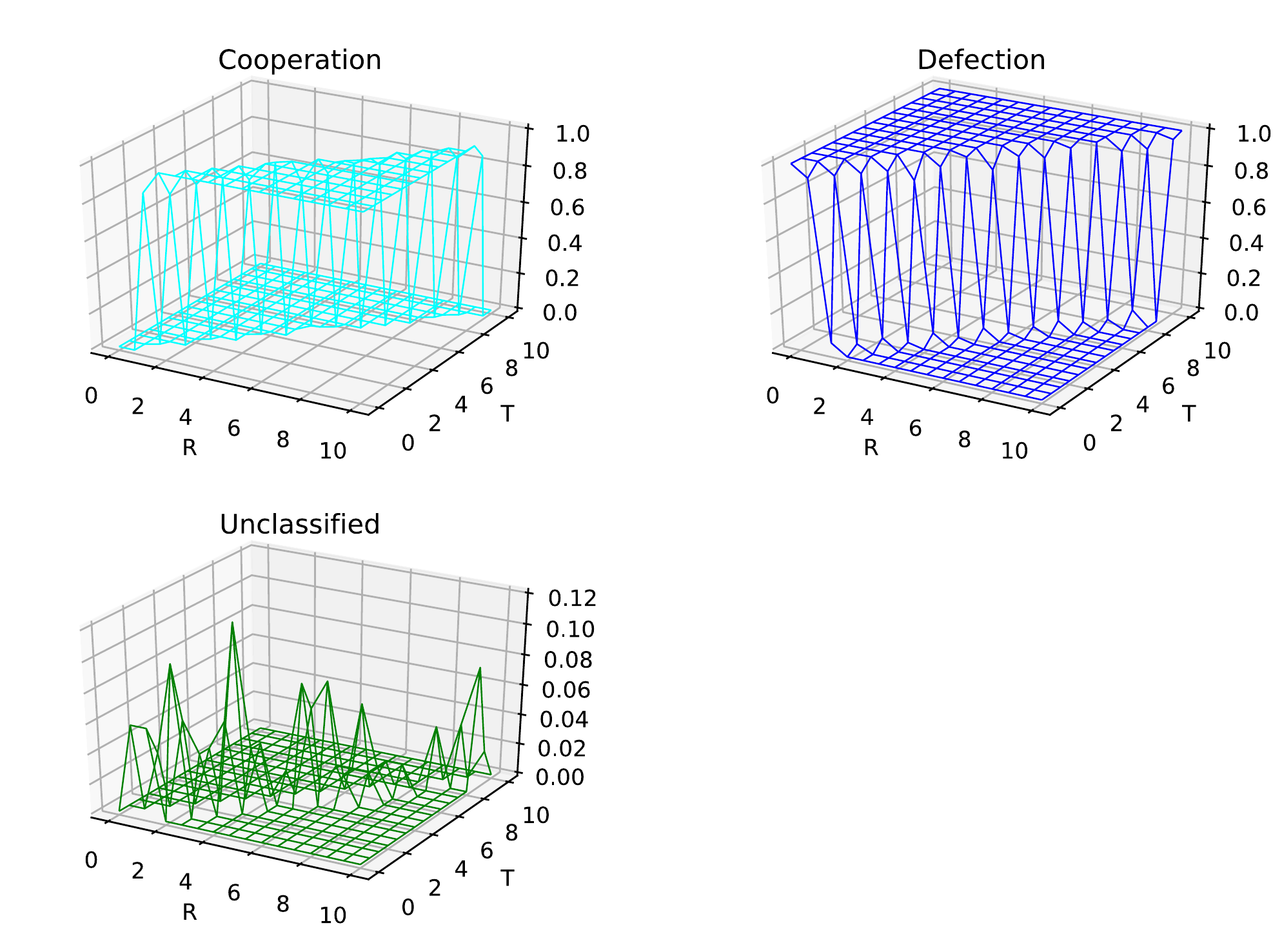}
    \caption{\textbf{3D Wire plots of the distribution of equilibria on varying $R$ and $T$ in the Prisoners' Dilemma using Wright-Fisher dynamics.} Simulations were run for 100 generations with results for each pair of $R$ $[0,10]$ and $T$ $[0,10]$ values averaged over 50 iterations.}
    \label{fig:wireplot}
\end{figure}

We can also produce 3D wire frame plots (\textbf{Figure \ref{fig:wireplot}}),  on varying 2 parameters of the game. Varying R and T as above, the only change in the code is to specify the type of graph to be created.

\begin{verbatim}
s = VariedGame(PrisonersDilemma, WrightFisher)
s.vary_2params(`R', (0, 10, 15), `T', (0, 10, 15),num_iterations=50,\ 
num_gens=100, graph=dict(type=`3d', lineArray=[(0, 10, 0, 10)]))
\end{verbatim}

The x and y axis correspond to the parameters being varied whereas the equilibrium proportion is plotted on the z axis.

\newpage
\bibliographystyle{plainnat}
\bibliography{SI.bib}